\documentclass[
 aip,
 reprint
]{revtex4-1}

\usepackage{upgreek}
\usepackage{float}
\usepackage{xcolor}
\definecolor{goodblue}{RGB}{13,71,160}

\usepackage{hyperref}
\hypersetup{
	colorlinks=true,
	linkcolor=goodblue,
	urlcolor=goodblue,
	citecolor=goodblue
}
\usepackage{amsmath}
\usepackage{amsthm}
\usepackage{amssymb}

\usepackage{graphicx}
\usepackage{dcolumn}
\usepackage{bm}

\usepackage[utf8]{inputenc}
\usepackage[T1]{fontenc}
\usepackage{mathptmx}
\usepackage{etoolbox}

\makeatletter
\def\@email#1#2{%
 \endgroup
 \patchcmd{\titleblock@produce}
  {\frontmatter@RRAPformat}
  {\frontmatter@RRAPformat{\produce@RRAP{*#1\href{mailto:#2}{#2}}}\frontmatter@RRAPformat}
  {}{}
}%
\makeatother

\begin{document}
\preprint{AIP/123-QED}

\title[High-Power Quantum-Limited Photodiode and Classical Laser Noise Squashing]{High-Power Quantum-Limited Photodiode and Classical Laser Noise Squashing}

\author{Vincent Dumont}
\author{Jiaxing Ma}
\author{Eamon Egan}
\author{Jack C.~Sankey}
\email{jack.sankey@mcgill.ca}
\email{vtdumont@gmail.com}

\affiliation{Department of Physics, McGill University, Montr\'{e}al, Qu\'{e}bec, Canada}
	
\date{\today}

\begin{abstract}
To benefit high-power interferometry and the creation of low-noise light sources, we develop a simple lead-compensated photodetector enabling quantum-limited readout from 0.3 mW to 10 mW and 10 k$\Omega$ transimpedance gain from  85 Hz - 35 MHz. Feeding the detector output back to an intensity modulator, we squash the classical amplitude noise of a commercial 1550 nm fiber laser to the shot noise limit over a bandwidth of 700 Hz - 200 kHz, observing no degradation to its (nominally $\sim 100$ Hz) linewidth.  

\end{abstract}

\maketitle

\section{Introduction}\label{sec:Intro}

All interferometric sensing and control experiments benefit from light sources and detectors operating at the shot noise limit. In systems employing a standard light source (even an ideal laser), shot noise places basic limits on readout and backaction noise  \cite{Aspelmeyer2014Cavity, Clerk2010Introduction}, while in more complex quantum control systems \cite{Rossi2018Measurement, magrini2021real, tebbenjohanns2021quantum}, sensing \cite{Degen2017Quantum}, and squeezing  \cite{wu1986generation, Safavi2013Squeezed, purdy2013strong, Brooks2012Non} experiments -- which can achieve sensitivities \emph{below} these limits \cite{grote2013first, Mason2019Continuous, Yu2020Quantum} -- attaining shot-noise-limited light represents an essential first step. Equally importantly (as is well-introduced in Ref.~\cite{grote2016high}), traditional room-temperature photodiodes capable of receiving power $P$ larger than a few milliwatts cannot achieve shot-noise-limited readout for three conspiring reasons: (1) photocurrent $I$ is always converted to voltage using a dissipative impedance $R$, (2) at temperature $T_R$, the ratio of this resistor's Johnson noise to the shot noise of the (mean) current $\bar{I}$ is $2k_B T_R/e\bar{I}R$ \cite{grote2016high}, and (3) the maximum operating voltages of low-noise electronics ($\sim$ 15 V) places a ceiling on the DC voltage $\bar{I}R$ across this resistor, placing a lower bound on this ratio. As is also discussed in Refs.~\cite{grote2016high, grote2007high}, if one is willing to shunt the DC with an inductor, one can use a larger value of $R$ within the targeted frequency range. In the interest of sensing acoustic frequencies relevant to gravitational waves, their design enabled shot-noise-limited readout with shunted photocurrents $\bar{I}$ up to 20 mA. The primary drawback of this approach is the inductor's inherent susceptibility to magnetic noise -- from both external fields \textit{and} its core -- requiring a mu-metal enclosure and specialized core materials.

In this work, we first present a comparatively simple and low-cost photodetector employing a lead compensating filter to reduce (but not eliminate!) the transimpedance gain at DC. Our initial design (optimized for 40 kHz optomechanical experiments \cite{Reinhardt2016Ultralow}), achieves a ``flat'' transimpedance gain of 10 k$\Omega$ from 85 Hz - 35 MHz (3 dB bandwidth) that drops to 1 k$\Omega$ at DC, providing a convenient diagnostic readout. Within the high-gain band, the detector is shot-noise-limited for $\bar{I} >$ 0.3 mA and can handle up to $\sim$10 mW incident light at 1550 nm wavelength. Importantly, this performance is achieved with standard electronic components and no magnetic shielding. While the present circuit serves our immediate purposes, it could be modified to further reduce the DC gain (further boosting the power handling, provided the diode itself can handle it), and / or trade bandwidth for increased in-band gain.\footnote{Please contact us for further discussion.}

We then demonstrate that feeding this diode's signal back to an intensity modulator allows us to squash the relative intensity noise (RIN) of a commercial 1550-nm fiber laser output, producing mW-scale output that is shot noise limited from 700 Hz - 200 kHz (3 dB), and within 5\% of the predicted shot noise over the majority of the band. Importantly, this feedback has no measurable effect on our laser's (nominally $\sim 100$ Hz linewidth) frequency noise spectrum.

To motivate the need for high-power photodiodes, we begin by reviewing amplitude noise squashing in Sec.~\ref{Sec:Overview_Squashing}, then present our design in Sec.~\ref{Sec:PD}, and finally test the performance of a noise squashing loop using it in Sec.~\ref{Sec:FeedbackNoiseSuppresion}.

\section{Intensity Noise Squashing Apparatus}\label{Sec:Overview_Squashing}

\begin{figure}[!htbp]
	\centering
     \includegraphics[width=0.95\columnwidth]{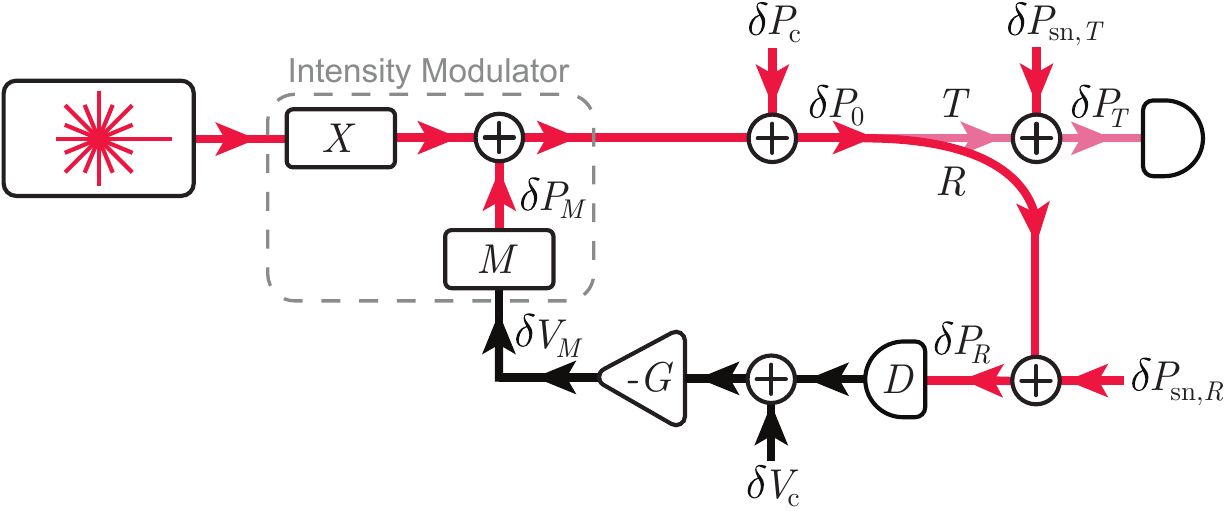}
	\caption{Classical noise squashing feedback loop with effective noise sources. Laser light passes through an intensity modulator with transfer functions (TF) $X$ transmitting input power to output power, and $M$ converting the control voltage $\delta V_M$ to added power $\delta P_M$. Classical amplitude noise from the laser and modulator are modeled as a subsequent single equivalent external source $\delta P_c$, producing total closed-loop noise $\delta P_0$. A fraction $T=0.07$ (0.57 mW) of the usable light is sent to the ``experiment'' diode (Thorlabs PDA-10CF), while a fraction $R=0.93$ (7.6 mW) is sent to our homemade ``feedback'' photodetector (TF $D$). Quantum (shot) noise is modeled as additional noise sources $\delta P_{\text{sn},T}$ and $\delta P_{\text{sn},R}$, arising from (uncorrelated) counting statistics at the diode inputs, $\delta P_T$ is the total output noise we hope to minimize, and $\delta P_R$ is the in-loop collected noise. The feedback diode output is filtered and amplified (all electronics lumped into a single TF $-G$) before feeding back to the intensity modulator. Photodiode, amplifier, and other electronic noises are also lumped into a single equivalent source $\delta V_\text{c}$ (dominated by photodiode noise).}
	\label{Fig1} 
\end{figure}

The key elements of a simple intensity noise squashing feedback scheme are shown in Fig.~\ref{Fig1}. In our case, a fiber laser (NKT Photonics Koheras Adjustik X15) sends 18 mW of 1550-nm light into a zero-chirp intensity modulator (iXblue MXAN-LN-10) DC-biased by a voltage-divided battery (lower noise than a power supply) to 90\% of its maximal output; we model this throughput with an overall constant transfer function $X$ [W/W] including all optical losses. The modulator's output is split (Thorlabs 10202A-90-APC) such that a fraction $T = 0.07$ (power $P_T= 0.57$ mW after all losses\footnote{The transmission $T$ and reflection $R$ coefficients relate to the measured currents (i.e., taking account each photodiode's responsivity), but we refer back to power for simplicity.}) is transmitted ``out of loop'' to an ``experiment'' photodetector (Thorlabs PDA10CF), while a fraction $R=0.93$ (power $P_R=7.6$  mW after all losses) is sent to the in-loop ``feedback'' photodetector (homemade using a Hamamatsu G12180-005A photodiode) transfer function $D$ [V/W]) discussed below. The resulting voltage is then fed through a proportional-integral (P-I) controller (Newport LB1005) and a 60 kHz single-pole low-pass filter (together having transfer function $-G$ [V/V]) back to the intensity modulator's RF control port, which is here modeled as a transfer function $M$ [W/V] that adds the (small) power fluctuations $\delta P_M$ to the original output.

For algebraic simplicity, we employ effective classical laser noise source $\delta P_c$ [W/$\sqrt{\text{Hz}}$] that combines the inherent classical noise from the laser and amplitude modulator (excluding feedback), and detection noise source $\delta V_c$ [V/$\sqrt{\text{Hz}}$] including all electronic noise of the diode and subsequent electronics. Note that one can separate the classical noise into different sources littered throughout the loop, but since they are all linearly related to each other via fixed transfer functions, they end up combining into one effective noise source; here we choose to keep electronic and laser noise separate as a conceptual aid. On the other hand, shot noise is modeled as frequency-independent spectra $\delta P_{\mathrm{sn}, T} = \sqrt{2 \bar{P}_T hc/\lambda} $ and  $\delta P_{\mathrm{sn}, R}=  \sqrt{2 \bar{P}_R hc/\lambda}$ [W/$\sqrt{\text{Hz}}$] that are separately added in quadrature (these noises are uncorrelated) to the laser power collected by each detector.

To self-consistently solve this loop, we first write the in-loop laser noise $\delta P_0$ as a sum of the inherent noise $\delta P_\mathrm{c}$ and the loop-modified copy of itself ($-[(\delta P_0 R + \delta P_{\text{sn},R})D+\delta V]GM$), then solve for $\delta P_0$, and finally convert this to the ``experiment'' noise [W/$\sqrt{\mathrm{Hz}}$]
\begin{align}\label{Eq:6-CNS-Fluctuation}
    \delta P_T &= \delta P_{\mathrm{sn},T} +  \frac{T}{1 + RDGM} \left[ \delta P_\mathrm{c}  - (GM) \delta V_c - (DGM) \delta P_{\mathrm{sn},R} \right].
\end{align}
Assuming all noises are uncorrelated, the transmitted noise power spectral density (PSD) [W$^2$/Hz] relative to the shot noise PSD becomes
\begin{align}
         \frac{S_{P_T}}{S_{P_{\mathrm{sn}, T}}}  &= 1  +  \frac{|T|^2}{|1 + RDGM|^2} \frac{S_{P_\mathrm{c}} + |GM|^2 S_{V_c} + |DGM|^2 S_{P_{\mathrm{sn}, R}}}{S_{P_{\mathrm{sn}, T}}}
\end{align}
using the convention $S_\alpha \equiv |\delta \alpha|^2$ for all symbols $\alpha$. To approach the shot noise limit, we hope to minimize the second term through careful loop design.

To gain further insight, suppose we have squashed the laser's classical noise ($S_{P_c} \ll |DGM|^2 S_{P_{\text{sn}},R}$) over a frequency range in which the loop gain is high ($|RDGM| \gg 1$). The above ratio then reduces to 
\begin{align}
         \frac{S_{P_T}}{S_{P_{\mathrm{sn}, T}}}  &\approx 1  +  \frac{|T|^2}{|R|^2}  \frac{  S_{P_{\mathrm{sn}, R}}  +  S_{V_c}/|D|^2
         }{S_{P_{\mathrm{sn}, T}}} \nonumber \\
         &\approx 1  +  \frac{|T|}{|R|}  \left[ 1 +  \frac{S_{V_c}} {S_{P_{\mathrm{sn}, R}}|D|^2} \right]\label{Eq-Expected-Noise-Reduction},
\end{align}
where in the last step we exploit the relationship $T/R = S_{P_{\mathrm{sn}, T}}/S_{P_{\mathrm{sn}, R}}$. Inspecting the last form, we see immediately that even an ideal feedback loop ($S_{V_c}=0$) will convert the in-loop shot noise to classical noise that cannot be squashed (the shot noise at each detector is uncorrelated). This simultaneously motivates the need for high in-loop power ($R\gg T$), and the correspondingly high-power, shot-noise-limited detector ($S_{V_c}\ll S_{P_{\mathrm{sn}, R}} |D|^2 $) discussed below.

\section{Photodetector Design}\label{Sec:PD}

\begin{figure}[!h]
	\centering
	\includegraphics[width=0.95\columnwidth]{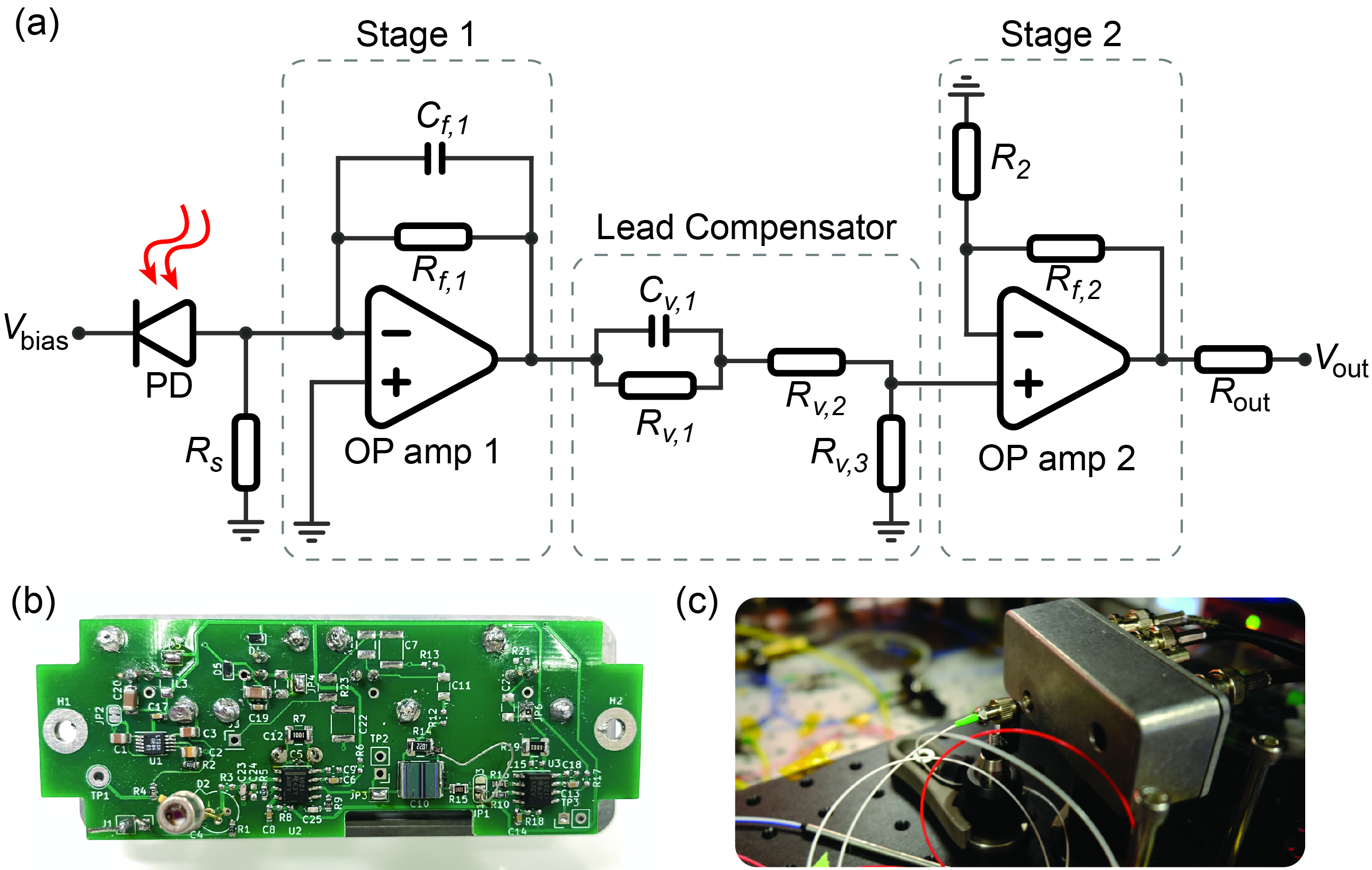}
	\caption{High-power, high-bandwidth, low-noise lead-compensated photodiode. (a) Simplified schematic of the photodiode circuit. A photodiode (Hamamatsu G12180-005A with sensitivity $\mathcal{S} = 1.1$ A/W   at wavelength $\lambda = 1.55$ $\upmu$m) biased at $V_\mathrm{bias} = + 10$ V sends current through an op-amp (THS4021) wired as a transimpendance amplifier (gain $G_1 = -R_{f,1} =   -1$ k$\Omega$), with a trimmer capacitor $C_{f,1} = 2-5$ pF used to tune the bandwidth ($1/2\pi R_{f, 1}C_{f,1} = 30-70$ MHz ) and stability of the circuit. A resistance $R_s =470$ $\Omega$ is also placed in parallel with the photodiode to improve the circuit's stability. To avoid saturating the second-stage op amp (THS4021) with the DC voltage resulting from the DC photocurrent, the DC gain is reduced by an interstage lead-compensating filter with $C_{v,1} = 10$ $\upmu$F, $R_{v,1} = 2.2$ k$\Omega$, $R_{v,2} = 100$ $\Omega$, and $R_{v,3} = 100$ $\Omega$, yielding attenuation $R_{v,3}/(R_{v,1} + R_{v,2} + R_{v,3}) = 0.042$ at low frequency ($\omega/2\pi\lesssim 10$ Hz) and $A_\mathrm{HF} = R_{v,3}/(R_{v,2} + R_{v,3}) = 0.5$ above $\sim$ 100 Hz. The second amplification stage (non-inverting amplifier with with $R_2 = 49$ $\Omega$, $R_{f,2} = 1000$ $\Omega$, and gain $G_2 = 1 + R_{f,2}/R_2 = 21.4$) ensures that the shot noise associated with 10 mA photocurrent is well above the input-noise ($\sim 10$ nV/$\sqrt{\mathrm{Hz}}$) on subsequent electronics. The  circuit output impedance is set by the resistance $R_\mathrm{out} =  50$ $\Omega$. Images below show (b) the circuit board and (c) assembled detector circuit during operation, shown for scale.  The full circuit is shown in Appendix \ref{Sec:App-PD Circuit}, and all source files are available upon request.}
	\label{Fig:2-PD-scheme} 
\end{figure}

To squash the classical noise on milliwatt-scale light, we require a feedback photodiode capable of shot-noise-limited readout at powers $\gtrsim 10$ mW. We employ a simple resistive-capacitive circuit with a lead-compensating filter as shown in Fig.~\ref{Fig:2-PD-scheme}(a); this provides a gain that is small enough at DC to handle high powers, but large enough over a bandwidth of interest to remain shot-noise limited.

\begin{figure}[!htbp]
	\centering
	\includegraphics[width=1\columnwidth]{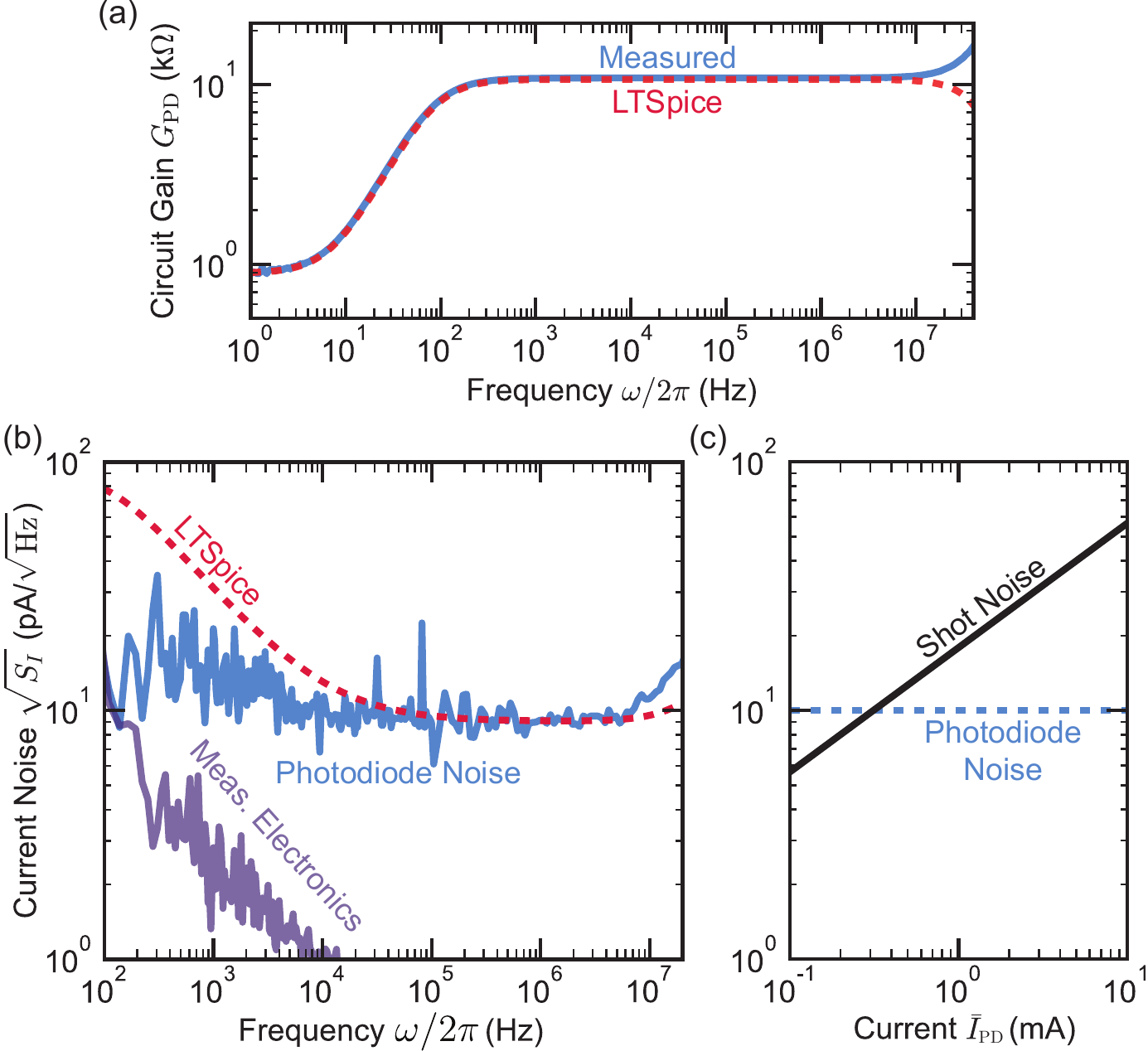}
	\caption{Amplifier circuit characterization. (a) Transimpedance gain measured (blue line) by injecting frequency-swept current through a diode-bypassing input resistor, and simulated (red dashed line) by LTSpice. (b) Input-referred current noise spectral density measured (blue curve) from the voltage noise output divided by the measured transimpedance gain in (a), with the purple curve showing the electronic noise after the amplifier output for reference. The red dashed line shows the noise as predicted with LTSpice.  (c) Comparison of mid-bandwidth noise (blue dashed line) and predicted shot noise (black line) for different values of photocurrent $\bar{I}_\mathrm{PD}$.  At current $\bar{I}_\mathrm{PD}  \gtrsim 300$ $\upmu$A (or laser power $ \bar{P}\gtrsim 270$ $\upmu$W), the photodiode will be shot noise limited. }
	\label{Fig:3-PD-Performance} 
\end{figure}

Our circuit is also tuned to achieve high bandwidth, with a gain of 1 k$\Omega$ at DC and 10 k$\Omega$ from 85 Hz to 35 MHz, as shown in Fig.~\ref{Fig:3-PD-Performance}(a). The actual circuit gain (blue curve) is measured by replacing the photodiode with a resistor of resistance $R_\mathrm{in}\approx 10$ k$\Omega$ and applying an oscillatory voltage $V_\mathrm{in}$ through it. This generates an oscillatory current $I_\mathrm{in}=V_\mathrm{in}/R_\mathrm{in}$ (relative to the op-amp's virtual ground), and we can then measure the board's output voltage $V_\mathrm{out}$, to estimate the photodetector gain  $G_{PD} \equiv V_\mathrm{out}/I_\mathrm{in}=R_\mathrm{in}V_\mathrm{out}/V_\mathrm{in}$. This agrees with the circuit simulation (LTSpice, red dashed curve) up to $\sim$ 10 MHz, where the actual first stage has a resonance due to stray reactance.  A photodiode with smaller area (i.e. lower capacitance) can help further stabilize this resonance if needed.

In our circuit, the most critical electronic noise sources to consider are those of the feedback resistor $R_{f,1}$, stability resistor $R_s$, and op-amp inputs of the first stage. For the chosen components, these three sources contribute almost equally (above 300 Hz, where the low frequency tails of the op amp's current and voltage noise decrease below the feedback resistor's Johnson noise). First, the feedback resistor Susumu HRG3216P-1001-D-T1 -- chosen for its negligible excess noise (beyond Johnson) at frequencies above $\sim$ 10 Hz \cite{seifert2009resistor} -- produces Johnson noise $\sqrt{S_V} = \sqrt{4 k_B T R_{f_1}} = 4.1$ nV/$\sqrt{\mathrm{Hz}}$ at the output of the first stage. Second, the Johnson noise of the stability resistor contributes $\sqrt{S_V} = \sqrt{4 k_B T R_{f_1}^2/R_s} =5.9$ nV/$\sqrt{\mathrm{Hz}}$ at the output of the first stage (due to voltage gain $R_{f,1}/R_s$). Third, the op amp's current noise contributes (a specified) $\sqrt{S_V} = G_1\times 1.2 \text{ }\mathrm{pA}/\sqrt{\mathrm{Hz}} = 1.2 \text{ nV}/\sqrt{\mathrm{Hz}}$ through the resistor with an additional (specified) voltage noise $\sqrt{S_V} = (1+R_{f,1}/R_s)\times 1.5 \text{ }\mathrm{nV}/\sqrt{\mathrm{Hz}} = 4.7 \text{ nV}/\sqrt{\mathrm{Hz}}$. Figure \ref{Fig:3-PD-Performance}(b) shows the input-referred current noise, measured (blue) by dividing the output voltage noise by the measured gain in (a), in agreement with summing the above noise sources in quadrature and dividing by the wideband transimpedance gain $G_1 = - 1$ k$\Omega$, which yields 8.7 pA/$\sqrt{\mathrm{Hz}}$. Importantly, the circuit has the expected noise floor over the targeted bandwidth. At low frequency, the actual noise is lower than that predicted by LTSpice (red), likely because the op-amp manufacturer's LTSpice noise model overestimates the actual amp's noise. At higher frequency, the actual noise is higher due to the circuit resonance. In Fig.~\ref{Fig:3-PD-Performance}(c) we compare the measured in-bandwidth photodiode noise (blue) to the expected shot noise for varied photocurrents. Notably, above 300 $\upmu$A ($\sim 270$ $\upmu$W laser light), the photodiode is shot-noise-limited, and at 10 mA ($\sim 9$ mW), the shot noise should be 5 times higher than circuit noise, in principle enabling classical noise squashing to within 6\% of shot noise (amplitude spectral density) with 90\% of the original light used for feedback (see Eq.~\ref{Eq-Expected-Noise-Reduction}).

\section{Feedback Noise Suppression}\label{Sec:FeedbackNoiseSuppresion}

\begin{figure}[!htbp]
	\centering
	\includegraphics[width=0.8\columnwidth]{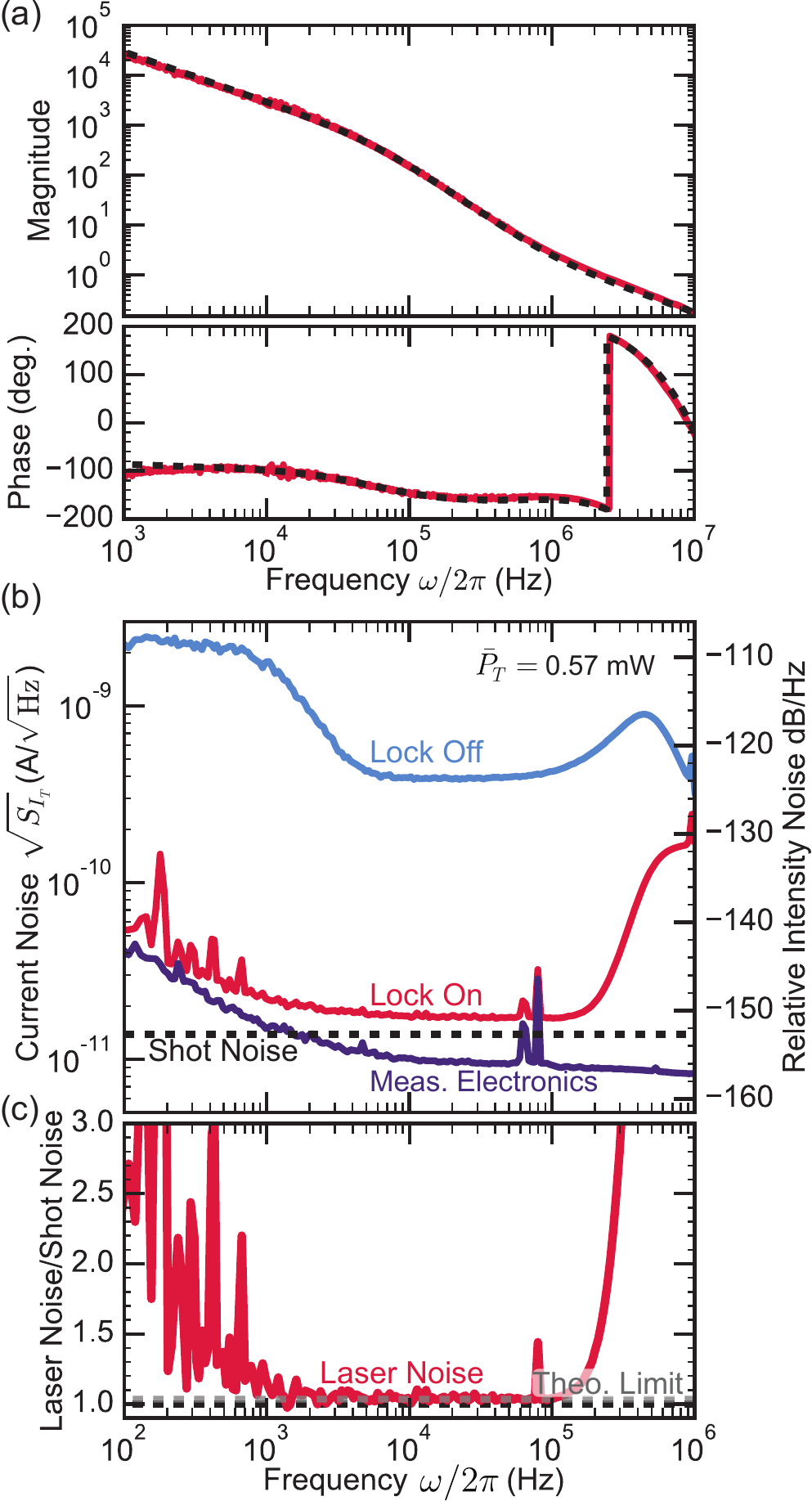}
	\caption{Feedback and noise squashing performance. (a) Magnitude (top) and phase (bottom panel) of the open-loop transfer function, estimated (red) as in Refs.~\cite{Reinhardt2017Simple,Janitz2017High}, and modeled (black dashes) as discussed in the main text. (b) Current amplitude spectral density (left $y$-axis) and relative intensity noise (right $y$-axis) at the ``experiment'' photodiode (Thorlabs PDA10CF) measured with feedback off (blue) and on (red). The purple data are the diode's electronic (readout) noise and the dashed black line is the expected shot noise level. (c) Ratio of laser noise to shot noise after subtracting electronic noise. The dashed black line indicates the shot noise limit, and the grey dashed line indicates the theoretical limit for an ideal feedback diode for the measured in-loop and out-of-loop photodiode currents (Eq.~\ref{Eq-Expected-Noise-Reduction}).}
	\label{Fig:4-Int_Noise_Performance} 
\end{figure}

We can insert this photodiode in the feedback loop of Fig.~\ref{Fig1} to squash classical laser noise. The subsequent electronics have a transfer function $-G$ (in Fig.~\ref{Fig1}) defined primarily by a proportional-integral (P-I) servo (Newfocus LB1005) with P-I corner frequency $\omega_\mathrm{PI}/2\pi = 1$ MHz, and a 60 kHz single-pole low-pass filter. As shown in Fig.~\ref{Fig:4-Int_Noise_Performance}(a), we estimate the open-loop transfer function (red data) of the entire loop by injecting a small modulation at the P-I servo's second input and measuring the closed-loop response through the error monitor (following Refs.~\cite{Reinhardt2017Simple, Janitz2017High}). This matches a theoretical calculation (black dashed line) for the chosen filters with loop delay $\tau_d = 79$ ns (consistent with $\sim$ 50 ns P-I controller delay and  $\sim$ 30 ns delay associated with optical/electronic cables  length) and an overall gain $2.7 \times 10^{-3}$ W/V. With these filters, the -180 degree phase-shift occurs at $\approx 2.5$ MHz, defining the feedback bandwidth.

Figure \ref{Fig:4-Int_Noise_Performance}(b) shows the experiment photodiode's input-referred current noise with the feedback off (blue) and on (red), along with the expected shot noise (black dashes) for the measured DC value $\bar{I}_T = 0.60$ mA (corresponding to collected power $\bar{P}_T \approx 0.57$ mW). The purple curve shows the detector's noise floor in the absence of light. The circuit squashes the classical intensity noise by a factor of 20 up to 200 kHz, above which the open-loop gain approaches unity with a phase that enhances noise. Note this squashing bandwidth can be widened by reducing the overall loop gain at the expense of less squashing in-band. As shown in Appendix \ref{Sec:App-PhaseNoise}, this squashing adds no appreciable frequency noise to the light.

Subtracting the detector's readout noise yields the actual laser noise, which is plotted in  Fig.~\ref{Fig:4-Int_Noise_Performance}  (bottom) relative to the shot noise. Importantly, the ``experimental'' light is dominated by shot noise over more than two decades of frequency (700 Hz - 200 kHz), and within 4 \% of the shot noise limit over the majority of the band. This agrees with the theoretical limit (grey dashed line) of 3.5\% for an ideal shot-noise-limited feedback diode at the observed DC value  $\bar{I}_R = 8.4$ mA ($\bar{P}_R \approx 7.6$ mW).

\section{Conclusion}\label{sec:Conclusion}
We have presented a low-cost, room-temperature, lead-compensated, shot-noise-limited photodetector with 35 MHz bandwidth that can handle up to 10 mW of laser light. The strategy, which can permit even higher power handling by reducing the DC gain, is nominally limited only by the power handling of the chosen photodiode. Alternatively, the circuit bandwidth might be improved by using a smaller-area photodiode, and other op amp compensation techniques (or the choice of a different op amp) might alleviate the need for the ``stability'' resistor that currently dominates the low-frequency noise. Removing this resistor could reduce the current noise to $\sim$ 4.5 pA/$\sqrt{\mathrm{Hz}}$, i.e. dominated by the Johnson noise of the feedback resistor.

We used this detector to squash the amplitude noise of a commercial fiber laser to the shot noise limit from 0.7 - 200 kHz at 0.57 mW. The bandwidth is limited by detector noise at low frequencies and the delay introduced by long fibers, cables, and our servo controller at high frequencies. The feedback circuit could be improved by, most importantly, reducing the bandwidth-limiting $\sim$ 80 ns delay  to $\sim 10$ ns by replacing the commercial servo with a fixed-gain, P-I filter (1-2 ns delay for typical integrated circuits) and shortening all fibers and cables. Additionally, we note that the choice of ``experimental'' power 0.57 mW is somewhat arbitrary. At lower power (by changing the optical splitter e.g.,), the total noise will be closer to shot noise than the 4\% margin demonstrated here, and at an experimental power of $\sim$ 10 mW the noise would be within a factor $\sqrt{2}$ (3 dB) above the quantum limit. Additionally, if one is interested in squashing classical laser intensity noise at a specific (potentially higher) frequency range, one could use a narrowband filter as was done in a different context in Ref.~\cite{parniak2021high}.

\section*{Funding}
 VD acknowledges financial support from FRQNT-B2 Scholarship and McGill Schulich Graduate Fellowship. JCS acknowledges relevant support from the Natural Sciences and Engineering Research Council of Canada (NSERC RGPIN 2018-05635), Canada Research Chairs (CRC 235060), Canadian foundation for Innovation (CFI 228130, 36423), Institut Transdisciplinaire d'Information Quantique (INTRIQ), and the Centre for the Physics of Materials (CPM) at McGill.

\begin{figure*}[!htbp]
	\centering
	\includegraphics[width=0.8\textwidth]{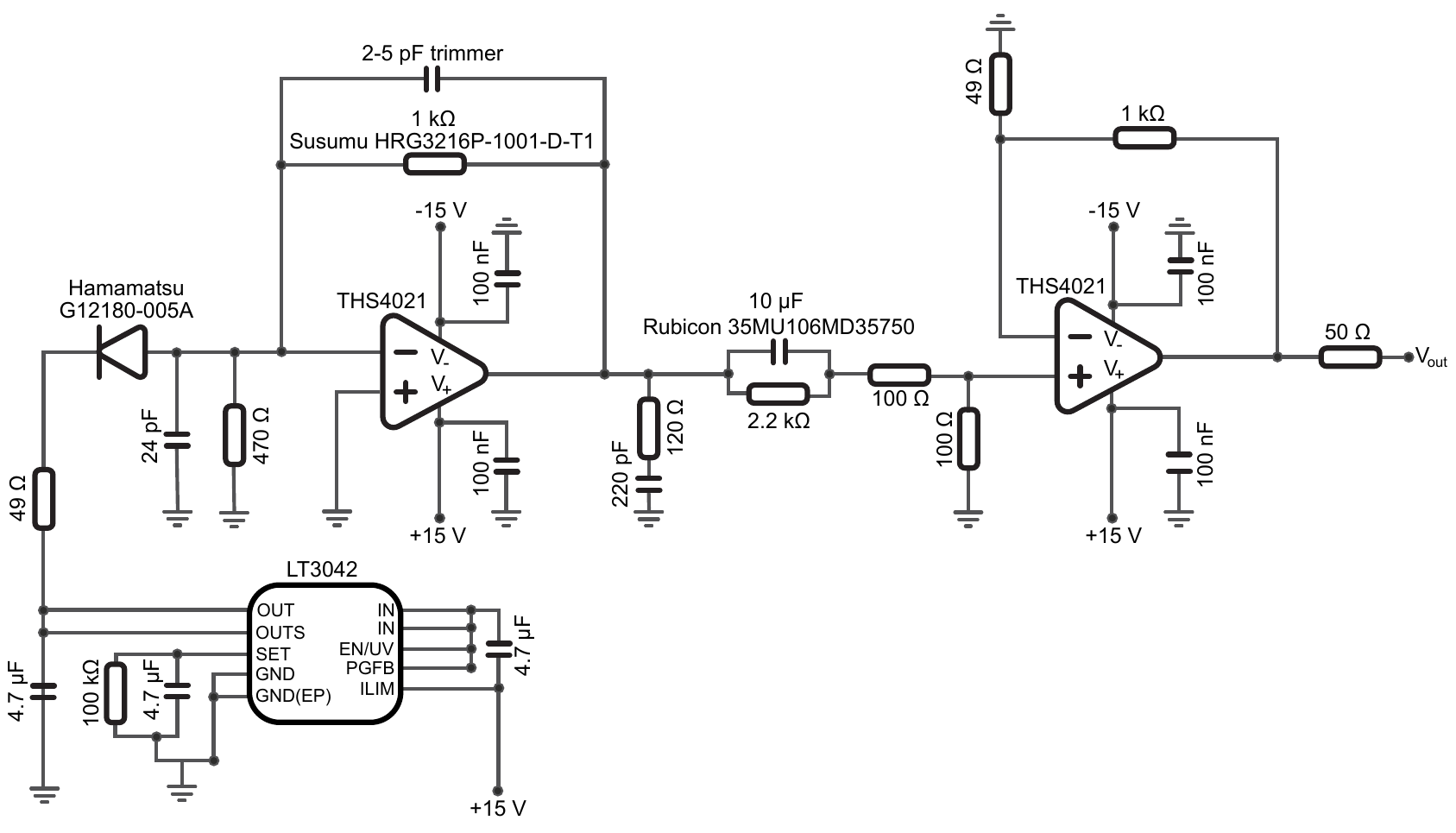}
 	\caption{Full schematic of the photodiode circuit. Part numbers are indicated when specific parts were used. }
	\label{Fig:App-Full} 
\end{figure*}

\section*{Data Availability Statement}
The data that support the findings of this study are available from the corresponding author upon reasonable request.

\section*{Acknowledgments}
We thank Simon Bernard, Thomas J. Clark, Hartmut Grote, Massimiliano Rossi, and Albert Schliesser  for fruitful discussions.

\section*{Disclosures}
The authors declare no conflicts of interest.

\appendix

\section{Full Photodiode Circuit}\label{Sec:App-PD Circuit}

The full circuit of the homemade photoddetector is shown in Fig.~\ref{Fig:App-Full}, and all source files are available upon request.

The photodiode is reverse biased at +10 V with an ultralow noise linear voltage regulator (LT3042). The photodiode (Hamamatsu G12180-005A) is chosen for its high quantum efficiency (0.88, which could be furthered increased by removing the protecting glass window) and its linear response within our power range (up to 10 mW). An additional 24 pF capacitance is placed in parallel with the photodiode to make the input node capacitance more predictable, and to allow testing and tuning without the photodiode present; a 470 $\Omega$ resistor is also added in parallel, which improves circuit stability by increasing noise gain (at the expense of greater noise). A $2-5$ pF trimmer  capacitor is placed in parallel with the 1 k$\Omega$ feedback resistor for fine-tuning the first stage compensation to achieve stability, while also limiting the stage's bandwidth. The feedback resistor (Susumu HRG3216P-1001-D-T1) was selected for its low excess (above Johnson) $1/f$ noise \cite{seifert2009resistor}. Another measure to improve stability consists of a 120 $\Omega$ resistor in series with a 220 pF capacitor shunting the first stage output to ground. Both (decompensated) op amps (THS4021), inspired by the design of Ref.~\cite{Mason2019Continuous}, are chosen for their large (2 GHz) gain-bandwidth product, low voltage (1.5 nV/$\sqrt{\mathrm{Hz}}$) and current (1.2 pA/$\sqrt{\mathrm{Hz}}$) noise levels, and moderately high maximum voltage supply handling ($\pm$16 V), to permit an elevated DC output voltage corresponding to the DC photocurrent. Other design considerations and functioning principles are presented in the main text (Sec.~\ref{Sec:PD}).

\section{Effect on Laser Phase Noise}\label{Sec:App-PhaseNoise}

\begin{figure}[!htbp]
	\centering
	\includegraphics[width=0.85\columnwidth]{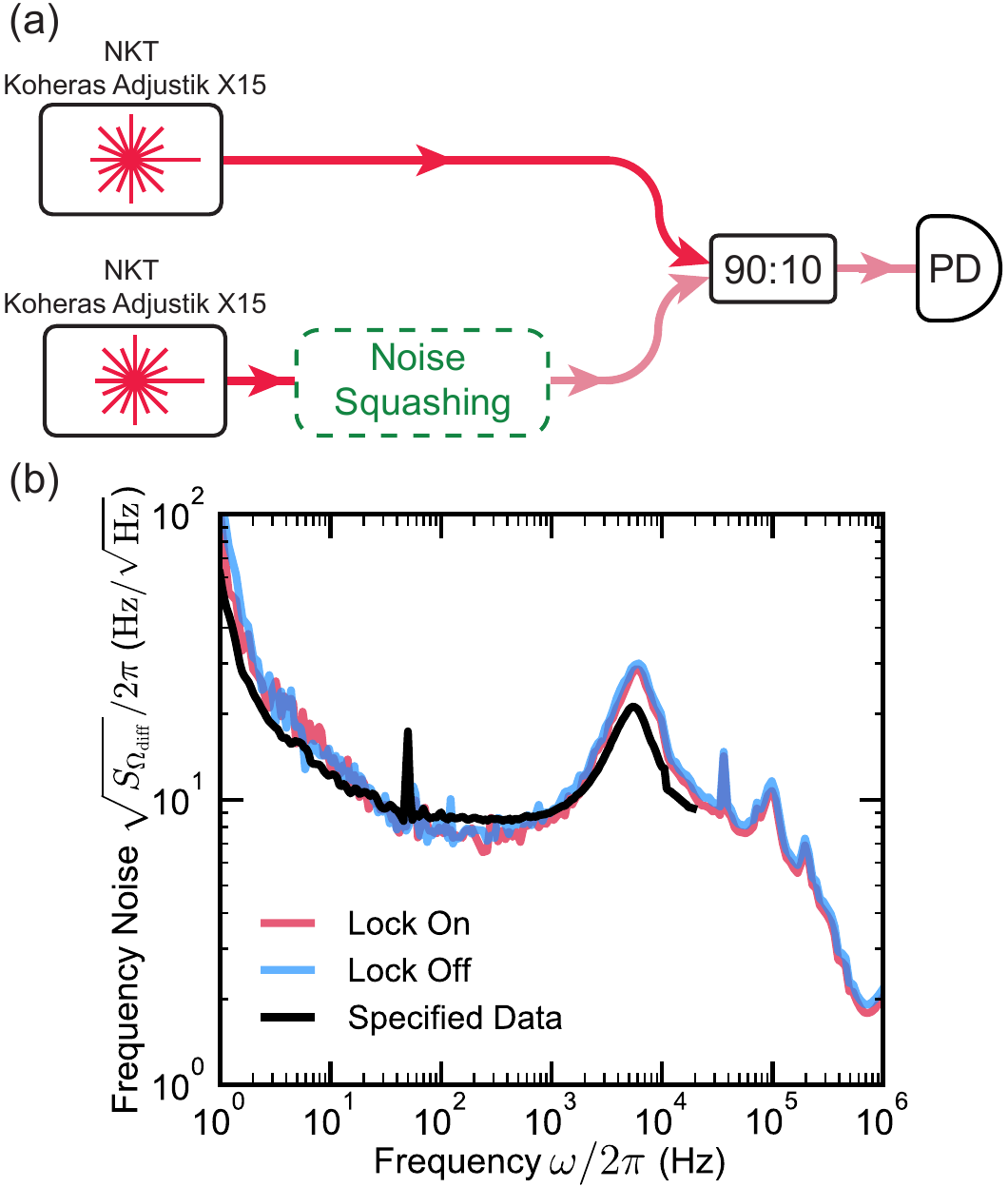}
 	\caption{Laser frequency noise with and without laser intensity noise feedback. (a) Illustration of the laser frequency noise measurement. Two lasers of the same model (NKT Koheras Adjustik X15) but slightly different frequencies $\omega_{\mathrm{in}, j}$, one of which which passes through the intensity feedback loop, are combined on a 90:10 splitter such as their powers are  approximately matched at the diode. The resulting beat note at difference frequencies $\Omega_\mathrm{diff} \equiv \omega_{\mathrm{in}, 2} - \omega_{\mathrm{in}, 1}$ varies with time due to frequency noise.  (b) Laser frequency noise amplitude spectral densitywhen the laser is locked (red), unlocked (blue). The manufacturer specification (summed for two laser) is shown in black. The small (systematic) offset between the two curves, e.g., above 1 kHz, are within observed drifts over time.}
	\label{Fig:6-IF-PSD-PN} 
\end{figure}

Here we present the measurement of our laser (NKT Photonics Koheras Adjustik X15) frequency noise, before and after the classical intensity noise squashing feedback circuit, showing no significant increase in frequency noise relative to our laser's nominal level $\sim 2\pi \times 10$ Hz$/\sqrt{\mathrm{Hz}}$. The measurement scheme, depicted in Fig.~\ref{Fig:6-IF-PSD-PN}(a), comprises the output of two lasers (same model) combined and collected by a single photodiode, the signal of which provides access to a frequency noise estimate.

To see how beating two beams at a photodiode reveals phase (or frequency) noise, we first write the electric field $E_j$ generated by each laser (index $j={1,2}$) after the optical splitter as
\begin{align}
    E_j = E_{0,j} \cos[\omega_{\mathrm{in}, j} t + \phi_j(t)] ,
\end{align}
with field amplitude $E_{0,j}$, angular frequency $\omega_{\mathrm{in}, j}$, and phase $\phi_j(t)$ encapsulating the phase noise we want to extract. The total field $E \equiv E_1 + E_2$ is effectively squared by the photodiode, which leads to a measured voltage $V \propto   E^2$:
\begin{align}\label{Eq:6-beating}
    V \sim  \frac{E_{0,1} E_{0,2}}{2}\cos[( \omega_{\mathrm{in}, 2} - \omega_{\mathrm{in}, 1})t +  \phi_2(t) - \phi_1(t) ],
\end{align}
where we have ignored terms rotating at $\sim 2 \omega_{\mathrm{in},j}$, as they are at extremely high frequencies and averaged away, and the DC term which is not relevant for our analysis.

For ``fast'' frequency fluctuations ($\omega/2\pi > 200$ Hz), we numerically demodulate this signal at the beat frequency $\Omega_\mathrm{diff} \equiv \omega_{\mathrm{in}, 2} - \omega_{\mathrm{in}, 1}$ (filtering away the term oscillating at $2\Omega_\mathrm{diff}$), thereby obtaining the laser phase difference $\phi_2(t) - \phi_1(t)$. From this, we directly compute the phase noise PSD $S_{\phi_\mathrm{diff}}$, from which we calculate the frequency noise PSD of the two combined lasers using $S_{\Omega_\mathrm{diff}}(\omega) = \omega^2 S_{\phi_\mathrm{diff}}(\omega)$. For ``slow'' frequency fluctuations ($\omega/2\pi < 200$ Hz), we simply use a phase-locked loop to track the beating signal as a function of time, from which we directly compute the frequency noise PSD. We present these combined measurements in Fig.~\ref{Fig:6-IF-PSD-PN}(b). Note that this method of interfering two different lasers yields the sum of the phase noise of each laser (assuming their noises are uncorrelated), but has the advantage of not requiring long fibers or feedback-stabilization of one interferometer arm's path length typical in self-intereference methods \cite{parniak2021high}. This measurement is done both while the intensity noise squashing feedback is off (blue) and on (red), with  the quadrature sum of the manufacturer's specified frequency noise of both lasers (black) shown for reference (and sanity check). All the curves show frequency noise performance similar to the laser manufacturer's specification, and the difference between them is consistent with observed drifts in the measurement (not seen to arise from the feedback). Thus, as seen  in Fig.~\ref{Fig:6-IF-PSD-PN}(b), this intensity noise squashing scheme does not add significant phase noise.

\addcontentsline{toc}{section}{References}
\bibliographystyle{Bibliography/bibstyle-jack}
\bibliography{Bibliography/errthing}

\end{document}